# A locking clamp that enables high thermal and vibrational stability for kinematic optical mounts

Maggie Kautz[1], Laird M. Close[2], Jared R. Males[2]

[1]College of Optical Sciences, University of Arizona, 1630 E University Blvd, Tucson, AZ 85719.
[2]Steward Observatory, University of Arizona, 933 N Cherry Ave, Tucson, AZ 85719.

## ABSTRACT

One of the main pursuits of the MagAO-X project is imaging planets around nearby stars with the direct detection method utilizing an extreme AO system and a coronagraph and a large telescope. The MagAO-X astronomical coronagraph will be implemented on the 6.5 meter Clay Magellan Telescope in Chile. The 22 mirrors in the system require a high level of mirror stability. Our goal is less than 1 microradian drift in tilt per mirror per one degree Celsius change in temperature. There are no commercial 2inch kinematic optical mounts that are truly "zero-drift" from 0-20C. Our solution to this problem was to develop a locking clamp to keep our optics stable and fulfill our specifications. After performing temperature variation and thermal shock testing, we conclude that this novel locking clamp significantly increases the thermal stability of stainless steel mounts by ~10x but still allows accurate microradian positioning of a mirror. A provisional patent (#62/632,544) has been obtained for this mount.

## 1.0 INTRODUCTION

Since the environment of MagAO-X is non-isothermal, it is prone to misalignment[1,2,3,4]. There are no commercial 2inch mounts with zero drift with temperature or commercial 2inch non-spring loaded kinematic mounts available today. We cannot use springs since they vibrate too much (and are too sensitive to temperature changes) to meet our specification. Our solution is to develop new super stable (1µrad/C) locking clamp prototype. The locking clamp is comprised of a modified commercial kinematic optical mount, split so the half of the mount that contains the tip/tilt linear actuators is what remains. The clamp is mounted so the linear actuators, or ball drivers, are set to be co-linear with the center of each pad on the mount holding the mirror. The contact points of the locking clamp and the pads keep the optic in a stable position when affronted by temperature and vibrational changes by compensating for any movement with radial pressure.

It is interesting to note that the CTE of the stainless steel TPI 100 adjusters is 10 ppm/C. Hence if there is ~10mm of driver past the bushing then one expects ~100nm/C of drive motion just from thermal considerations alone which changes tilt in both the X and Y directions. Moreover, this change in tilt is compounded by the spring constant itself also decreasing leading to change in the gravity induced sag, and the mount further tilting in pitch (Y direction).

However, the linear actuators of our novel locking clamp should be able to halt these motions. On the microscopic level the steel/steel interface locks the mirror plane and optic's tilt in place to the less than 20nm/C level. The reason we cannot do much better than ~20nm/C may be due to the fact that alloy steel ball bearings have surface roughness of ~10 nm rms with peaks as high as 50 nm. Hence, these materials are intrinsically "rough" at these levels and we may be becoming dominated by microphysics. In this paper we describe this clamp and how well it performed in numerous lab tests of stability as the temperature of the lab (method #1) and then just the temperature of the mount itself (method #2) were changed.

## 2.0 METHODS

### 2.1 Method #1: Changing the temperature of the optics lab

Initially, we placed two optical kinematic mounts with 2inch flat mirrors across from each other on an optical table approximately 3m apart. A laser source was reflected between them three times, creating a total beam path of 24.703m (see Figs. 1 and 2). We taped a piece of paper to a vertical flat surface next to the mirror on the left side of the table. An airy pattern from the final reflection was projected on the paper. I would outline the initial location of the airy pattern. In order to vary room temperature, we would turn the Air Conditioning in the room on and off. A thermocouple was attached to the table to track lab temperature. As I noticed the temperature dropping or increasing by more than one degree I would outline the new location of the airy pattern. The translations of the airy pattern were plotted against the temperature changes. This experiment was carried out with the both mirrors unlocked (no clamping), and then repeated with both mirrors locked. The formula for mirror tilt as a function of spot motion is as follows:

$$\delta = \frac{S_{target}}{18D} \cdot 1000000$$

$\delta = mirror\ tilt\ (\mu rad)$

$s_{target} = spot\ motion\ (mm)$

$D = 24703\ mm$

There were several issues with this initial method. The thermocouple was measuring the temperature of the table rather than the temperature of the mount itself. We could not rule out the warping of the table significantly influencing the movement of the airy pattern, so we needed to remove the table's temperature sensitivity as a variable.

### 2.2 Method #2: Thermally shocking the mount itself

Our second method tried to account for some of the issues that arose with our first method. An aluminum bowl was filled with boiling water and placed into an open cardboard box with a hole on the side. Several copper strips were stacked together to make a 1.375in wide, 0.06in thick strip and were screwed between the post and the kinematic mount holding the 2inch mirror and placed under the aluminum bowl. This caused a heat shock directly on the mount. A piece of graph paper was next to the mirror on the other side of the optical table so the final airy pattern's drift could be tracked against temperature change. The thermal couple measured the temperature of the mount directly. The tests were repeated with mounts locked (clamped) and unlocked (no clamp) several times.

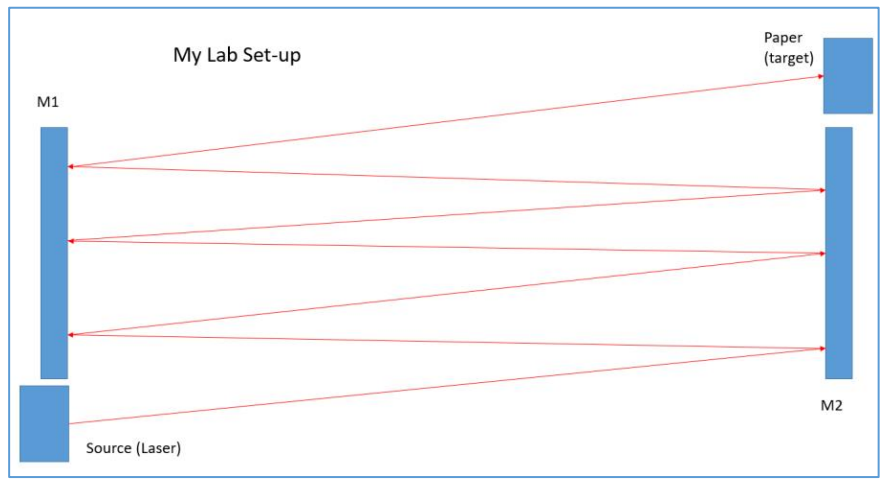

Figure 1. A schematic of initial lab layout, the distance from source to target was 24.703m.

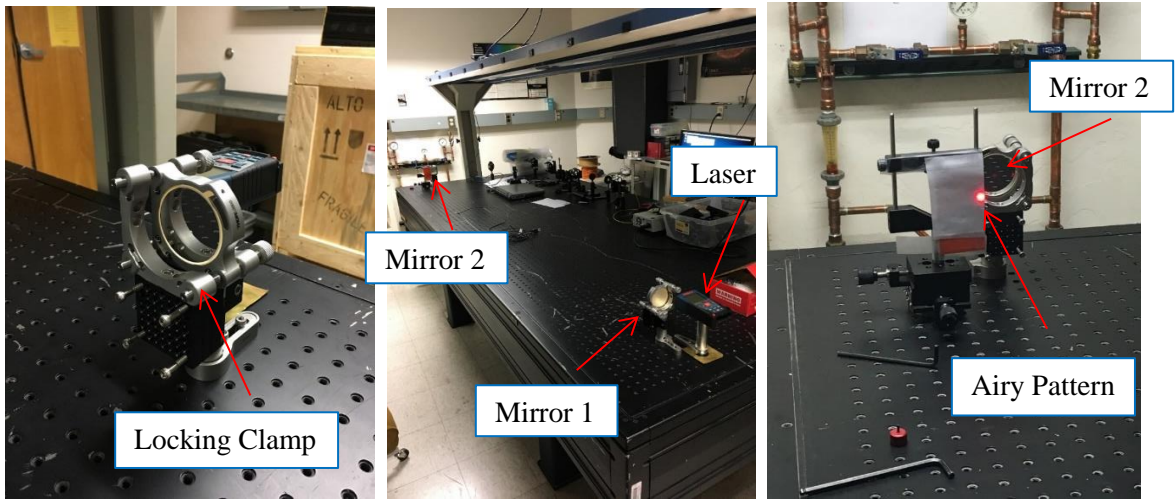

Figure 2. Photos of initial lab set-up

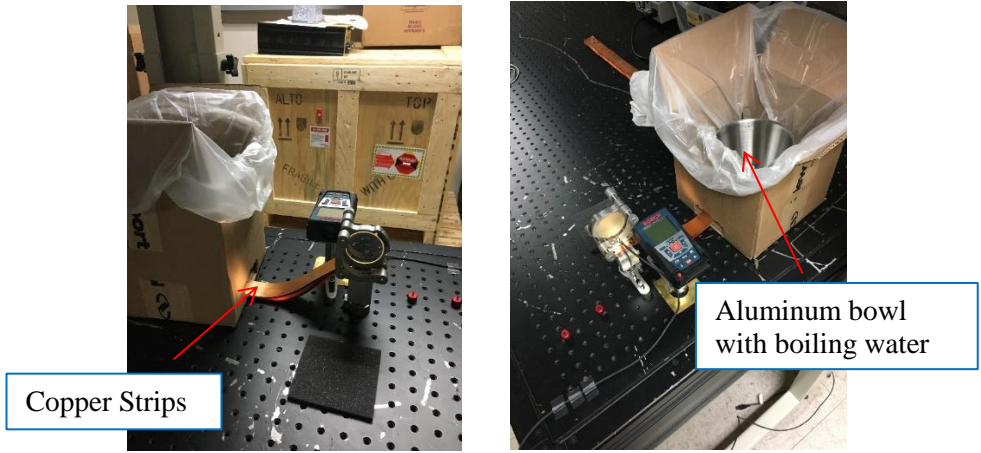

Figure 3. Photos of copper strip "direct thermal shock" (Method #2)

# 3.0 SPLIT RING POST CLAMP

The locking clamp is mounted on a custom designed steel split ring post clamp. The prototype split ring was manufactured in aluminum but the final MagAO-X clamp is fabricated out of 100% stainless steel to maintain (and match) the low CTE and low distortion of the system[3]. It also can provide mounting fiducials and alignment targets for quick and easy off-axis parabolic (OAP) alignment[3]. It also provides a safety feature (a Teflon strip) that prevents the escape of an OAP from the mirror cell in the unlikely case of a shipping mishap.

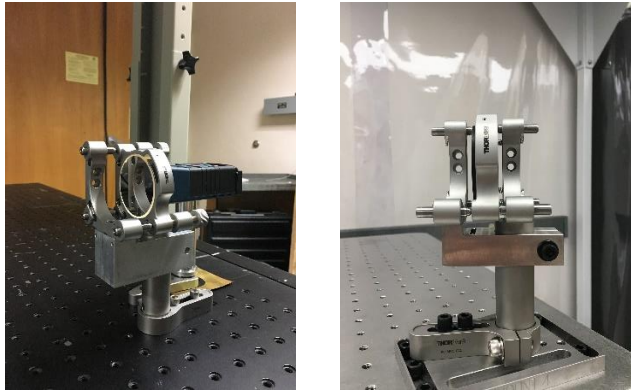

Figure 4. *(left)* Photos of the prototype and final *(right)* MagAO-X locking clamp and split ring system[3]

# 4.0 RESULTS

The results from method #1 indicated that each mount, when unlocked, had approximately 27.7 µrad tilt per degree C whereas once locked (clamps engaged) the mount had approximately 7.6 µrad tilt for every degree C (this was all in the Y direction (pitch; gravity direction), no significant X (yaw) motion detected. In Fig. 5 one can see the slope of the lines is measuring the change in pitch of two mounts against the change in temperature. The slopes are divided in half for our results since there are two mirrors accounting for the change in pitch in this experiment.

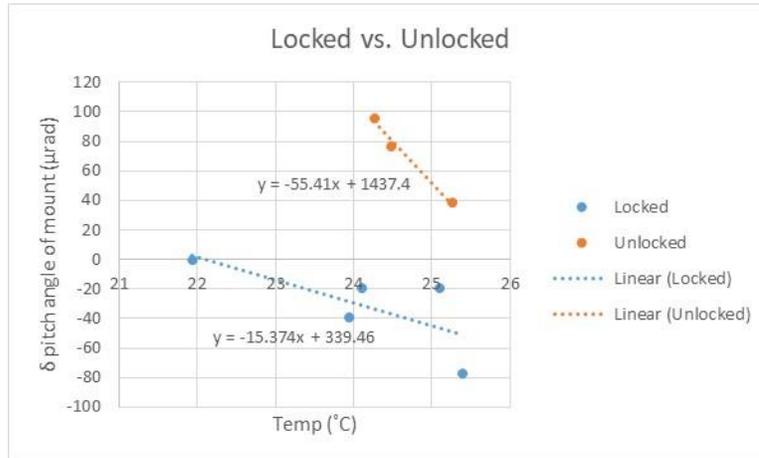

Figure 5. Method #1 Data

Nevertheless, a significant part of this motion is due to the optical table itself warping with temperature. So, this is an underestimate of the true stability of the clamps. To fix this we removed the effect of the table with direct thermal shock tests (method #2) – which just directly heated the mount itself. Fig. 6 shows how the location of the final airy pattern was tracked with a camera and its movement was measured on graph paper. There was no significant motion in the X direction. But in the Y direction (down w.r.t. gravity) we found movement to be 6±1 µrad/C unlocked (see Fig. 6) and 0.3-0.4 µrad/C in the locked configuration (See Fig. 7). We conclude that our custom clamping mount has >10 times better thermal stability (<0.6 µrad/C) than the best stainless steel mounts commercially available today.

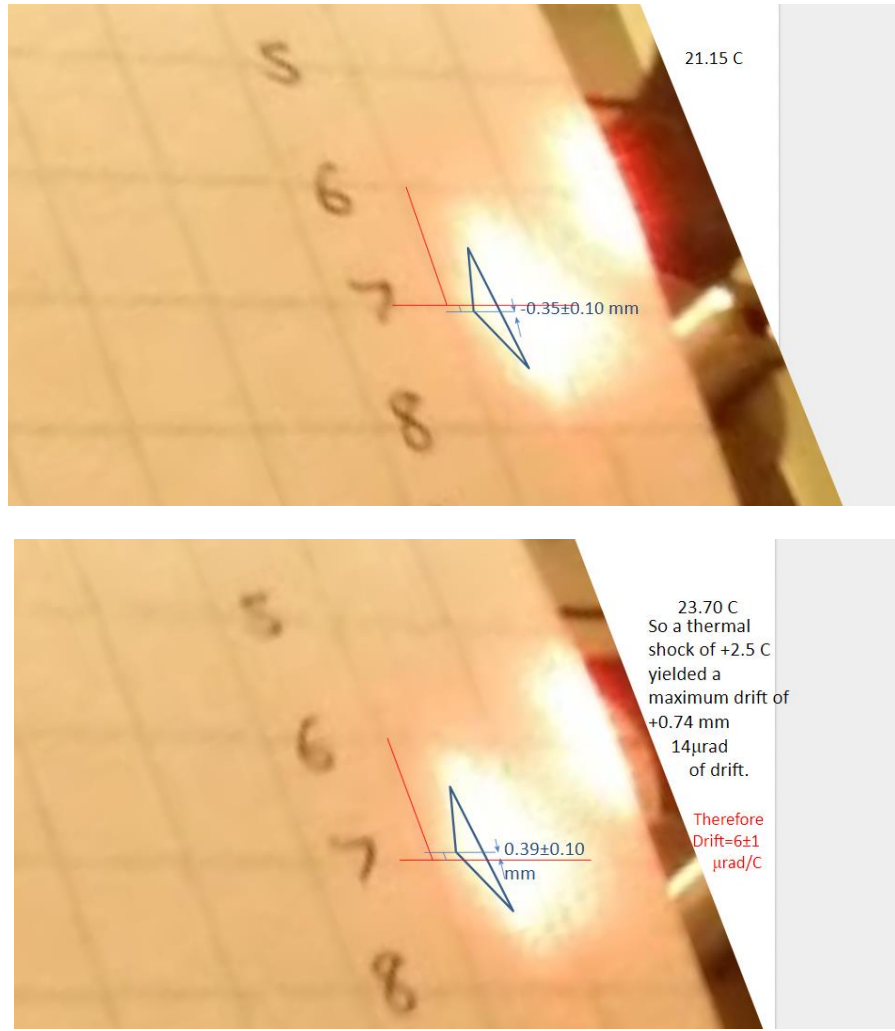

Figure 6. Photos of airy pattern (saturated white light). We used a simple matched filter PSF movement analysis to track the motion of the PSF. These images here are for the **unlocked** case. Without the locking clamp the stainless steel mounts have 6±1 µrad/C – which is worse than the MagAO-X spec.

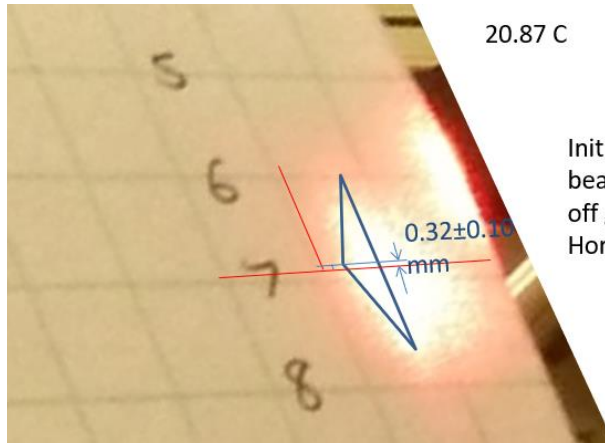

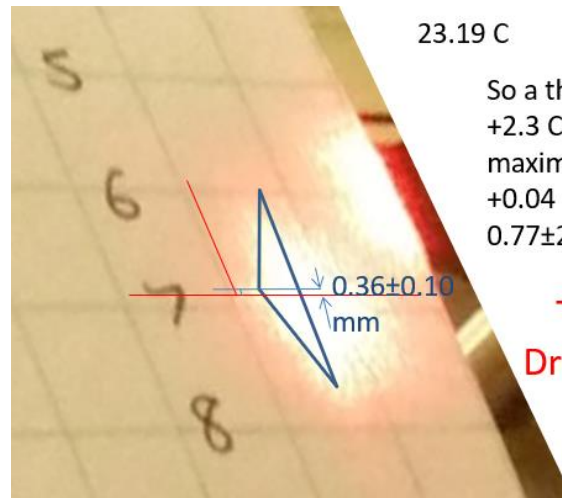

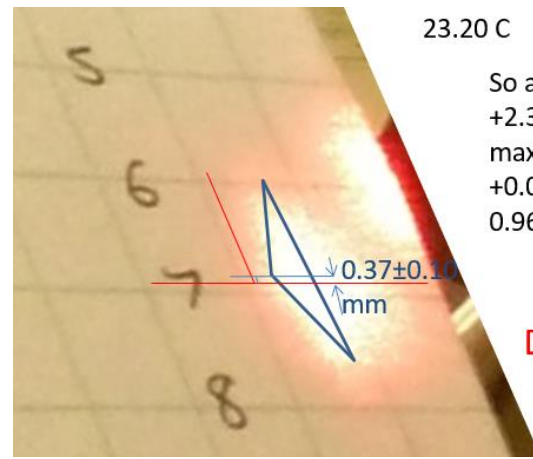

Figure 7. Photos of airy pattern (saturated white light). These images here are for the **locked** case. With the locking clamp the stainless steel mounts have 0.3-0.4 µrad/C – which meets the MagAO-X spec.

# 5.0 CONCLUSIONS

The achieved stability of our locking clamps (~0.6 μrad/C) suggests that for our long focal length (vertex focal length=621mm) OAPs (see Close et al. 2018[2]; these proc for the full optical mechanical MagAO-X design) there should only be 0.4μm/C of decenter per OAP in our novel clamped mount. Hence, over a delta temperature change of 10C (max change during a night at Magellan) we expect only 4μm of decenter/mirror. Over the 22 mirrors using the correct distance between each mirror[2] the cumulative decenter effect would be <45μm in the worst case of all the errors adding together after a 10 C change in instrument temperature. Detailed Zemax decenter analysis showed no effect of a 100 μm decenter of the optical system[2]. So, we conclude that there is unlikely to be a serious impact on the optical performance thanks to the stability of the clamps. It is also worth noting that the commercial mounts (without our clamps) had ~6μrad/C, hence they would have ~450 μm of decenter – which unacceptability high and would require an expensive manual re-alignment of MagAO-X when the temperature changed during the night by more than a few degrees C. So, in conclusion, these novel clamps enable us to both: 1) align MagAO-X; and 2) keep it aligned – even if the temperature changes by 10 C during the night.